\documentstyle[aps,prd,preprint,epsf]{revtex}
\tighten

\begin{document}
\draft

\title{Dynamics of perturbations of rotating black holes}
\author{William Krivan${}^{(1,2)}$, Pablo Laguna${}^{(1)}$,
Philippos Papadopoulos${}^{(1,3)}$, and Nils Andersson${}^{(4)}$}

\address{
${}^{(1)}$ Department of Astronomy \& Astrophysics and \\
Center for Gravitational Physics \& Geometry\\
Penn State University, University Park, PA 16802, USA}

\address{
${}^{(2)}$ Institut f\"ur Astronomie und Astrophysik\\ Universit\"at
T\"ubingen, D-72076 T\"ubingen, Germany}

\address{
${}^{(3)}$ Max-Planck Institut f\"ur Gravitationsphysik\\
D-14473 Potsdam, Germany}

\address{
${}^{(3)}$ Department of Physics\\
Washington University, St Louis, MO 63130, USA}

\date{\today}

\maketitle

\begin{abstract}
We present a numerical study of the time evolution of perturbations of
rotating black holes.  The solutions are obtained by integrating the
Teukolsky equation written as a first-order in time, coupled
system of equations, in a form that explicitly captures its hyperbolic
structure.  We address the numerical difficulties of solving the
equation in its original form. We follow the propagation of generic
initial data through the burst, quasinormal ringing and power-law
tail phases. In particular, we calculate the effects
due to the rotation of the black hole on the scattering of incident
gravitational wave pulses. These results may help explain how           
the angular momentum of the black hole affects the gravitational waves
that are generated during the
final stages of black hole coalescence.
\end{abstract}

\section{Introduction}
\label{intro}

Numerical relativity, {\em i.e.} the numerical construction of
solutions to the Einstein equations, 
is a rapidly advancing field. Reliable multi-dimensional 
simulations of astrophysically
relevant scenarios should become possible in the near future.
The systematic exploration of the
non-linear content of Einstein's theory is of 
major importance for this accelerated progress.  In the
pursuit of numerically solving non-linear systems, {\em intrinsic}
accuracy tests, such as convergence, help establish the reliability of
the solution and its proximity to the continuum limit.  However,
recurrent experiences suggest that tests, {\em extrinsic} to the
discretization process, are also valuable tools for assessing the
reliability of complicated numerical algorithms. One versatile tool,
providing a number of such extrinsic tests, is the investigation of the
linear regime of non-linear systems. In the appropriate limit,
numerical solutions of non-linear systems should agree with the
considerably simpler perturbative solutions. This testing approach is
bound to have fundamental importance to numerical relativity as the field
progresses.

Perturbative methods enjoy renewed attention in their own right.  The
emphasis now is on the application of the standard perturbative
methodology to particular realistic systems that are thought to be of
interest to the emerging field of gravitational wave
astronomy. Perhaps the most important one among those systems is the
collision of inspiraling black hole binaries.  The key premise of the
perturbative approach to black hole collisions is that, during the
last stages of the coalescence (the close limit), the spacetime can be
accurately approximated as that of a single perturbed black hole.  A
particular subclass of such problems is the study of black hole
collisions as perturbations of Schwarzschild spacetimes; these are
situations (e.g. head-on collisions) for which the end product of the
collision is a slowly or non-rotating black hole. Mathematically they
are described by the gauge invariant Zerilli function
\cite{zerilli}. The numerical solution of the Zerilli equation does not
pose difficulties and is routinely employed in estimating aspects of
gravitational wave physics in the context of black holes
\cite{Price}.  The generalization of this framework to rotating black
holes is physically relevant; it is likely that, during the inspiral
collision of black holes, the system will not be able to radiate all of its
angular momentum and will leave behind a single, rotating black hole.
The work in this paper constitutes an important
step towards the goal of generalizing the close-limit treatment of
black hole collisions  to that of perturbations about a single, rotating black
hole.  The general framework to achieve this goal requires: (A) the
construction of appropriate initial data describing two orbiting black
holes, (B) the subtraction of the Kerr background from the initial
data, to read off the initial perturbations, and (C) the evolution of these
perturbations.  In this paper, we concentrate on the last point,
namely the evolution of gravitational perturbations of the Kerr
spacetimes.

Previously, we investigated the dynamics of scalar fields in the
background of rotating black holes \cite{PaperI} (from here on
referred to as Paper I). We considered both slowly and
rapidly rotating black holes.  For slowly rotating black holes the
background geometry can be treated as a perturbed Schwarzschild
spacetime, with the
angular momentum per unit mass playing the role of a perturbative
parameter.  The study was centered on the late-time dynamics of the
scalar field and showed that, for rotating black holes of arbitrary
angular momentum, the late-time dynamics is dominated by the lowest
allowed multipole ($l=0$).  Here we
undertake the development of a numerical scheme for the
study of gravitational perturbations of Kerr black holes in the time
domain.  The main objective of this work is to follow the propagation
of generic initial data through the burst, quasinormal ringing and
power-law tail phases of the evolution.  In particular, we are
interested in investigating the effects of black hole rotation
on the scattering of incident gravitational wave pulses.  A
characterization of such effects provides an indication of the role
that rotation plays on signals produced during the final stages of
black hole coalescence.

\section{Perturbing Kerr black holes}

At first instance, a direct derivation of the equations governing
the perturbations of Kerr spacetimes is to consider perturbations of
the metric. This path, however, leads to gauge-dependent formulations.
A theoretically attractive alternative is to
examine {\em curvature} perturbations.  Using the
Newman-Penrose formalism, Teukolsky \cite{teuk72,teuk73}
derived a master equation
governing not only gravitational perturbations (spin weight $s = \pm 2$) but
scalar, two-component neutrino and electromagnetic fields as well.
In Boyer-Lindquist coordinates and with
the use of the Kinnersley null tetrad \cite{kinnersley}, this master
evolution equation reads
\begin{eqnarray}
\label{teuk0}
&&
-\left[\frac{(r^2 + a^2)^2 }{\Delta}-a^2\sin^2\theta\right]
         \partial_{tt}\Psi
-\frac{4 M a r}{\Delta}
         \partial_{t\phi}\Psi
- 2s\left[r-\frac{M(r^2-a^2)}{\Delta}+ia\cos\theta\right]
         \partial_t\Psi
\nonumber\\  &&
+\,\Delta^{-s}\partial_r\left(\Delta^{s+1}\partial_r\Psi\right)
+\frac{1}{\sin\theta}\partial_\theta
\left(\sin\theta\partial_\theta\Psi\right)
+\left[\frac{1}{\sin^2\theta}-\frac{a^2}{\Delta}\right]
         \partial_{\phi\phi}\Psi
\\  &&
+\, 2s \left[\frac{a (r-M)}{\Delta} + \frac{i \cos\theta}{\sin^2\theta}
\right] \partial_\phi\Psi
- \left(s^2 \cot^2\theta - s \right) \Psi = 0, \nonumber
\end{eqnarray}
where $M$ is the mass of the black hole, $a$ its angular momentum per
unit mass, and $\Delta\equiv r^2-2Mr+a^2$. For gravitational perturbations,
the function $\Psi$ is given
in terms of the Weyl tensor tetrad components $\psi_0$ and $\psi_{4}$.
That is, $\Psi = \psi_0$ for $s = +2$ and
$\Psi = \rho^{-4} \psi_{4}$ for $s=-2$, with
$\rho = - 1/(r - i a \cos \theta) $.

The Teukolsky equation~(\ref{teuk0}) reduces to the Bardeen-Press
equation \cite{bardeen} in the limiting case of non-rotating black
holes. 
For the case $s=0$, it yields the equation for a scalar wave propagating in
a Kerr background,  a system which we studied numerically in Paper I.

A result of great importance for perturbation theory was the discovery
by Teukolsky that, when viewed in the frequency domain,
Eq.\ (\ref{teuk0}) is separable in the $r$ and $\theta$
coordinates
(separation of the azimuthal angular dependence
is always possible).
To our knowledge, most of the work on the dynamics of perturbations of
Kerr spacetimes has been performed in the frequency domain  (or 
under the assumption of a harmonic
time dependence).  This has certainly been the case for studying
quasinormal mode frequencies
\cite{detweiler1980,leaver,seidel,kostas}, 
wave scattering \cite{Matzner}, and the motion of test particles
in the Kerr background \cite{sasaki,kojima1,kojima2}. 

Here we are interested in the time integration of physical initial 
data, possibly from the inspiral collision of binary black holes. 
The data may be analytic approximate solutions to the linearized 
constraints, or numerical solutions of the non-linear constraints.

In principle, one can Fourier transform the initial data and perform 
the evolution of such data for each frequency. Afterwards, the data 
are transformed back to the time domain if needed. 
From the computational point of view, however, this approach is 
far more expensive than it is to solve the problem in the time domain.
For several reasons the number of frequencies that one has to 
consider is orders of magnitude higher than the number of angular 
components required to resolve the $\theta$-direction: While the
angular directions are bounded, the time direction is only bounded
from below by the initial data surface. Focussing on the 
investigation of quasinormal modes (QNM) and power-law tails, one would first 
of all expect that one needs
quite a fine resolution near the $\omega=0$ point in order to be able to
correctly resolve the tails. Similarly, the resolution of the QNM
would be very sensitive to the spacing in frequencies. 
With this in mind, we have chosen to solve the Teukolsky
equation in $t,\,r,$ and $\theta$ coordinates.

The resulting 2+1 evolution equation is a hyperbolic, linear equation
which is quite amenable to numerical treatment, provided suitable
coordinates, variables and boundary conditions are chosen.
The major numerical difficulty in solving the Teukolsky equation in
the time domain arises from the term linear in $s$, involving the
first-order time derivative.
Depending on the relative sign between the coefficients of
$\partial_{t}\Psi$ and $\partial_{tt}\Psi$, one may view
$\partial_{t}\Psi$ either as a damping term (when the signs of both
coefficients agree) or an anti-damping term (otherwise).
Without the factor $2s$,
the real part of the coefficient of $\partial_{t}\Psi$ reads
\begin{equation}
\label{damping}
{\cal C}(r) =  \frac{M(r^2-a^2)}{\Delta}-r.
\end{equation}
In the physically allowed range $[r_+,\infty)$, where
$r_+ \equiv M + \sqrt{M^2-a^2}$ represents the event horizon,
the function $\cal C$ is monotonic in $r$, with $\lim_{r\to r_+}{ \cal C} =
\infty$
and $\lim_{r\to \infty}{ \cal C} = -\infty$. Therefore, there exists
a point $r_c$ such that ${\cal C}(r<r_c) > 0$
and ${\cal C}(r>r_c) < 0$. On the other hand, the
coefficient of $\partial_{tt}\Psi$ is always negative.
Consequently, for $s>0$, the term $\partial_{t}\Psi$ acts as a
(anti-)damping term for $(r<r_c)\, r>r_c$.
The situation reverses for $s<0$.
Of course, there is no real damping or anti-damping, in terms of
energy balance. The precise polynomial dependence of the coefficient(s)
is affecting, among other things, the required asymptotic fall-off
behavior of ingoing and outgoing radiation.
However, the above consideration suggests that, in numerical
work, the presence of such derivative terms could lead to
exponentially growing modes.
Those unphysical modes are not part of the family of solutions
to the Teukolsky equation. Indeed, in our first attempts of
solving the Teukolsky equation, such growing modes were
present in our simulations. These modes were clearly of numerical origin.
The suppression of these instabilities proved to be
an interesting and challenging
exercise in the construction of numerical algorithms.

The two key factors in successfully solving the Teukolsky equations were:
first, to rewrite equation (\ref{teuk0}) in a form that explicitly
exhibits the radial characteristic  directions, and second, to carefully
select the evolution field and its asymptotic behavior.

A successful numerical evolution of the Teukolsky equation was achieved by
considering  the $s=-2$ case. This restriction
does not diminish in any way the range of initial data that can be
evolved. In retrospect, the choice of $\psi_4$ as the evolution
field is natural since the radiation content at infinity is most 
succinctly described in terms of that field. 
Yet in our investigation this choice was dictated by the need to 
obtain numerically acceptable asymptotic limits near the horizon.
In order to reconstruct the perturbation of the 
metric coefficients from the solution of the Teukolsky equation, 
however, one would have to evolve the equation for $s=2$ as well. 
In our approach, i.e.\ evolving $\psi_4$, it is possible to recover 
the metric perturbations only at infinity. In some sense this restricts 
the information that we get from our calculations. Local information 
about the perturbed spacetime (about geodesics etc) is not available.

For a given spin weight $s$, outgoing waves correspond to solutions to 
Eq.\ (\ref{teuk0}) with the limiting behaviour \cite{teuk73}
\begin{eqnarray}
\lim_{r^* \to +\infty}{|\Psi_{s}|} & \sim & 1 / r^{2s+1} \\
\lim_{r^* \to -\infty}{|\Psi_{s}|} & \sim & 1
\label{asymp-out}
\end{eqnarray}
at spatial infinity and the event horizon, respectively. Meanwhile, 
ingoing waves behave as
\begin{eqnarray}
\lim_{r^* \to +\infty}{|\Psi_{s}|} & \sim & 1 / r \\
\lim_{r^* \to -\infty}{|\Psi_{s}|} & \sim & \Delta^{-s} \; .
\label{asymp}
\end{eqnarray}
Here we have used the Kerr tortoise coordinate $r^*$, which is defined by
\begin{equation}
\label{kerrtort}
dr^* = \frac{r^2+a^2}{\Delta} dr \, .
\end{equation}

The convenient properties of the $s=-2$ choice can be verified by
looking at the asymptotic form of propagating waves near the horizon
($r^* \to -\infty$). For $s=-2$, the solutions
are bounded for any direction of propagation; in contrast, for
$s=2$ the ingoing solution diverges as the horizon is approached
($\Delta \to 0$). The asymptotic behavior for $s=-2$ at $r^* \to +\infty$
can be subsequently fixed
by requiring that outgoing waves have
asymptotically bound amplitude also in this limit. 
For the Teukolsky function $\Psi_{-2}$,
this is achieved through a rescaling by an appropriate function of
$r$.  A convenient and simple choice is $r^3$, a factor that is regular at
the horizon.

Regarding the choice of spatial coordinates,
we use the Kerr tortoise
coordinate $r^*$ defined by Eq.\ (\ref{kerrtort}).
As azimuthal coordinate, we use
the Kerr $\tilde\phi$ coordinate instead of the Boyer-Lindquist coordinate
$\phi$.
The coordinate transformation to $\tilde\phi$ is defined by
\begin{equation}
d\tilde\phi \equiv d\phi + \frac{a}{\Delta} dr  \; .
\end{equation}
The azimuthal coordinate $\tilde\phi$ was previously used for the
scalar field evolution
in order to ameliorate coordinate pathologies near the
horizon. A discussion of those pathologies and their precise
manifestation in the slow rotation limit is given in Paper I.

\section{Solving the Teukolsky equation}

We can now introduce the following Ansatz for the solution to the
Teukolsky equation:
\begin{equation}
\Psi(t,r^*,\theta,\tilde\phi) \equiv e^{i m
\tilde\phi}\,r^3\,\Phi(t,r^*,\theta)\; .
\end{equation}
It follows that the Teukolsky equation for
$\Phi(t,r^*,\theta)$ has a structure similar to that of
Eq.\ (\ref{teuk0}), and therefore the same analysis of
 the damping or anti-damping nature of first-order
time derivative terms applies.
After a series of unsuccessful numerical experiments with this
second-order in time and space equation for $\Phi$,
we found that numerical instabilities were suppressed by
introducing an auxiliary field $\Pi$ that converts the Teukolsky
equation to a coupled set of first-order equations in space and time.
This can be accomplished
by defining
\begin{mathletters}
\begin{eqnarray}
\Pi &\equiv& \partial_t{\Phi} + b \, \partial_{r^*}\Phi \; , \\
b & \equiv &
\frac { {r}^{2}+{a}^{2}}
      { \Sigma} \; , \\
\Sigma^2 &\equiv &  (r^2+a^2)^2-a^2\,\Delta\,\sin^2\theta
\; .
\label{pi_eq}
\end{eqnarray}
\end{mathletters}
The resulting form of the Teukolsky equation has the following structure:
\begin{equation}
\partial_t \mbox{\boldmath{$u$}} + \mbox{\boldmath{$M$}}
\partial_{r^*} \mbox{\boldmath{$u$}}
+  \mbox{\boldmath{$L$}}\mbox{\boldmath{$u$}}
+  \mbox{\boldmath{$A$}}\mbox{\boldmath{$u$}}  = 0,
\label{new_teu}
\end{equation}
where $ \mbox{\boldmath{$u$}} \equiv
\lbrace \Phi_R, \Phi_I,
        \Pi_R,  \Pi_I \rbrace$ is the solution vector with
(I) and (R) labeling the imaginary and real parts, respectively.
The coefficient matrices {\boldmath{$M$}} and {\boldmath{$A$}} are
given by:
\begin{equation}
\mbox{\boldmath{$M$}} \equiv \left(\matrix{
                    b  &   0   &  0     &  0 \cr
                    0  &   b   &  0     &  0 \cr
                    m_{31}  &   m_{32}  & -b  &  0 \cr
                    -m_{32}  &   m_{31} &  0  & -b \cr
                }\right) \; ,
\label{m_matrix}
\end{equation}
and
\begin{equation}
\mbox{\boldmath{$A$}} \equiv \left(\matrix{
                    0  &   0   &  -1  &  0 \cr
                    0  &   0   &  0  &  -1 \cr
                    a_{31} & a_{32} & a_{33} & a_{34} \cr
                    -a_{32} & a_{31} & -a_{34} & a_{33} \cr
                }\right) \; ,
\label{a_matrix}
\end{equation}
respectively. Finally,  {\boldmath{$L$}} is a matrix operator
that contains all the angular derivatives and has the following
non-vanishing elements:
\begin{equation}
 \mbox{\boldmath{$L$}} \equiv \left(\matrix{
                    0  &   0   &  0  &  0 \cr
                    0  &   0   &  0  &  0 \cr
                    l_{31}  &   0   &  0  &  0 \cr
                    0  &   l_{31}   &  0  &  0 \cr
                }\right).
\label{l_matrix}
\end{equation}
The coefficients in the above matrices are given explicitly in
Appendix A.

The first-order system given by Eq.\ (\ref{new_teu}) is hyperbolic
in the radial direction 
since the eigenvalues of {\boldmath{$M$}} are real and there is a
complete set of linearly independent eigenvectors.
Diagonalization of the matrix {\boldmath{$M$}}
and construction of evolution schemes based on the eigenfields
generated stable evolutions. However, it turned out that
this last step was not necessary. Stable evolutions were also achieved using
a modified Lax-Wendroff method (see, for example, \cite{numrec})
applied to Eq.\ (\ref{new_teu}) when this equation was rewritten as:
\begin{equation}
\partial_t \mbox{\boldmath{$u$}} + \mbox{\boldmath{$D$}}
\partial_{r^*} \mbox{\boldmath{$u$}}
=  \mbox{\boldmath{$S$}}\; ,
\label{new_teu2}
\end{equation}
with {\boldmath{$D$}} = diag($b,b,-b,-b$) and
$\mbox{\boldmath{$S$}} = -(\mbox{\boldmath{$M$}} - \mbox{\boldmath{$D$}})
\partial_{r^*}\mbox{\boldmath{$u$}}
- \mbox{\boldmath{$L$}}\mbox{\boldmath{$u$}} 
- \mbox{\boldmath{$A$}}\mbox{\boldmath{$u$}}$.

Equation (\ref{new_teu2}) is discretized on a uniform two-dimensional polar
grid. 
Typically we used a computational domain of size $-100M \le r_i^* \le 500M$
and $0 \le \theta_j \le \pi$ with $0 \leq i \leq 8000$ and 
$0 \leq j \leq 32$.
The Lax-Wendroff updating of field values $\mbox{\boldmath{$u$}}^n_i$
given on a timestep $t_n$ consisted of two steps.
In the first step, intermediate field values
$\mbox{\boldmath{$u$}}^{n+1/2}_{i+1/2}$ are obtained from 
\begin{equation}
\mbox{\boldmath{$u$}}^{n+1/2}_{i+1/2} = 
\frac{1}{2} \left( \mbox{\boldmath{$u$}}^{n}_{i+1}
                  +\mbox{\boldmath{$u$}}^{n}_{i}\right)
- \frac{\delta t}{2}\,
\left[\frac{1}{\delta r^*} \mbox{\boldmath{$D$}}^{n}_{i+1/2}
  \left(\mbox{\boldmath{$u$}}^{n}_{i+1}
                  -\mbox{\boldmath{$u$}}^{n}_{i}\right)
- \mbox{\boldmath{$S$}}^{n}_{i+1/2} \right] \; ,
\end{equation}
where we have omitted the angular indices $j$ for clarity. 
The fields are $j$-centered in the angular direction
with angular derivatives approximated by the appropriate
second-order centered expressions. Radial derivatives in
$\mbox{\boldmath{$S$}}^{n}_{i+1/2}$ are approximated by centered
difference expressions of field values at $i$ and $i+1$. Similarly,
algebraic terms in 
$\mbox{\boldmath{$D$}}^{n}_{i+1/2}$ and $\mbox{\boldmath{$S$}}^{n}_{i+1/2}$
are obtained by averaging over the values at $i$ and $i+1$. 
The second and final step is a Staggered
leapfrog step:
\begin{equation}
\mbox{\boldmath{$u$}}^{n+1}_{i} = 
\mbox{\boldmath{$u$}}^{n}_{i}
- \delta t\, \left[\frac{1}{\delta r^*} \mbox{\boldmath{$D$}}^{n+1/2}_{i}
  \left(\mbox{\boldmath{$u$}}^{n+1/2}_{i+1/2}
                  -\mbox{\boldmath{$u$}}^{n+1/2}_{i-1/2}\right)
- \mbox{\boldmath{$S$}}^{n+1/2}_{i} \right] \, .
\end{equation}
Here the centered difference expressions and averages for
$\mbox{\boldmath{$S$}}^{n+1/2}_{i}$ and $\mbox{\boldmath{$D$}}^{n+1/2}_{i}$ are 
taken from values at $\mbox{\boldmath{$u$}}^{n+1/2}_{i+1/2}$
and $\mbox{\boldmath{$u$}}^{n+1/2}_{i-1/2}$. 
Examination of radial and angular characteristic directions at
arbitrary locations shows that the appropriate CFL condition is
$ \delta t \le \mbox{max}\{\,\delta r^*, \, 5\, \delta\theta\, \}$,
where $\delta t$ is the evolution timestep.

Boundary conditions are required at the horizon, at the rotation axis of the
black hole and at radial infinity.
The boundary conditions near the horizon are $\Phi = \Pi = 0$, as
follows
immediately from the asymptotic behavior of
ingoing waves (see Eq.(\ref{asymp})). The distinction of ingoing and
outgoing waves near the Kerr event horizon requires a careful
definition because of the rotational dragging of inertial frames 
\cite{teuk73}. However, in practice 
 it clearly does not
matter if we set both propagation modes to zero.
At the axis, only a condition on $\Phi$ is required.
Depending on the behavior of the angular eigenfunction belonging 
to the azimuthal number $m$ near the axis,
one imposes either $\Phi = 0$ or $\partial_\theta \Phi = 0$ \cite{Matzner}.
Finally, the boundary conditions at infinity are much harder to impose when
using a finite radial grid. Approximate schemes allow the transmission
of waves across the boundary, but the implied truncation of the
equations coefficients interferes strongly with physical features of
the evolution such as quasinormal ringing and tails. A clean solution
can be achieved via ``Cauchy-characteristic matching'' (see
\cite{ekgI,ekgII} and references therein). However, this approach will not be
pursued here. For the problems of immediate interest the
computational
domain can be made sufficiently large that errors generated at the
outer boundary will not affect the results. 
Consequently, we set $\Phi$ and $\Pi$ arbitrarily to zero at the
outer boundary, and our results are computed only from information inside
the maximal Cauchy development of the initial data surface.

The stability of the code was verified with long-time evolutions,
of the order of $1000M$. Its accuracy was tested using
standard convergence tests. The code was found to be second
order convergent for evolution times $t <50M$. The convergence rate
degrades as the total evolution time is increased, but is
consistently above $1.3$.
A note on the nature of the unstable modes discussed previously
is of some relevance here.
Linear systems with constant coefficients may exhibit numerical
instabilities if the eigenvalues of the update matrix have modulus
larger than unity. This phenomenon is easily analyzed with the
so-called Von Neumann analysis of the difference equation.  A
similar analysis can be used in equations with variable coefficients,
via local stability analysis, but the method becomes far less
conclusive in that case. For the Teukolsky equation, a number of
simple, locally stable, discretizations turned out to be unstable for
late times.  The instabilities depended on the numerical
discretization lengths, with increased resolution generally leading to
slower growth rates for the unphysical modes. Yet only impractically
high resolutions would suppress the instabilities for the long evolution
times required for  the applications discussed here.

\section{BURSTS, QUASINORMAL MODES AND POWER-LAW TAILS}

The initial data for
gravitational  perturbations of
rotating, asymptotically flat spacetimes can be broadly
divided into two classes: Waves coming in from infinity
and then scattering off the black hole background, and waves
emanating from the black hole as a result of an external
excitation of the black hole. The latter case is perhaps the
most interesting since it includes the description of the final stages of the
binary black hole coalescence within the close-limit approximation.

In general, the evolution of both types
of initial data consists of three stages, as seen
by an observer located  away from the hole.  During the first stage,
the observed signal depends on the structure of the initial pulse and
its reflection by the curvature potential  (burst phase). This
phase is followed by the exponentially decaying quasinormal ringing of
the black hole (quasinormal phase). In the last stage, the wave
slowly dies off as a power-law tail (tail phase).  The precise
manifestation of the last two phases is dictated by the subtle
interference of the respective amplitudes. The complex frequencies of
the QNMs for various values of $a$ are well known
\cite{detweiler1980,leaver,seidel,kostas}  and we can use them
as benchmarks of the present code's accuracy.  Similarly, the
late-time tail phenomenon is mathematically well understood in
spherically symmetric backgrounds, where tails have been predicted and
verified \cite{price72,gpp94a}. For rotating black holes, the
background is no longer spherically symmetric, but numerical results
for scalar waves in Kerr backgrounds suggest that power-law tails will
also be present for spin-2 fields \cite{PaperI}. Furthermore, it has
recently
been argued that these tails should take the same form as for
non-rotating black holes \cite{Nils}. 

We have used
the numerical algorithm described in the previous section
to obtain the time evolution of generic perturbations impinging
on the rotating black hole. For all the results presented here,
we have used ingoing initial data with compact support for $\Phi_R$ and
have set $ \Phi_I = 0 $.
All the evolutions 
clearly showed the exponentially damped oscillations of the
QNMs and the subsequent late-time power-law tail. We will now discuss
each of these phases in somewhat more detail.

\subsection{Power-law tails}

In Paper I we showed that the late time evolution of scalar
fields in the background of rotating black holes is qualitatively
similar to the non-rotating case. Here we extend this result to
the physically more interesting evolution of a spin-2 field.

Starting with initial data $\propto~_{-2}Y_2^m$ (where $\ _s Y_l^m$
denotes a spin-weighted spherical harmonic), we studied the
late time regime for $a/M=0,\, 0.25,\, 0.5,\, 0.75,\, 0.9$ and
$m=0,\, \pm 1, \pm 2$. In all cases,
we found a power-law tail behavior $|\Phi| \propto t^{-\mu}$.
For a nonrotating black hole it is known that the power-law
exponent should be $\mu = 2l+3$ \cite{gpp94a,Nils}. Indeed,
for $a=0$ we could
reproduce the theoretical value $\mu=7$ with an
accuracy of $10\%$. Moreover, our calculations show that
the exponents governing the behavior for
$a \ne 0$ remain similar to the Schwarzschild value. 
These results are summarized in 
Table \ref{tailvalues}. We conclude that
the power-law tail behavior is basically determined by 
the dominant asymptotic form of the potential, namely its 
``Schwarzschild" part. That this would be the case has recently
been suggested by analytic arguments \cite{Nils}. Analytic
results (for non-rotating black holes) can also explain the main
discrepancy between our numerically determined power-law exponents
and the theoretical values. The field will be governed
by a pure power law $t^{-(2l+3)}$ only at very late times. In a
somewhat earlier regime ``higher order'' terms will also be
significant
\cite{Nils}. The presence of such higher order terms tend to 
increase the value of a numerically extracted power-law exponent.

The results discussed above correspond to initial data that are dominated
by the lowest allowed multipole $l$ for the particular value of $m$ 
that is being considered ($l\ge |m|$). 
When this is not the case, we find that different  
multipoles become mixed in the
evolution. To investigate this, we 
started from an initial angular distribution 
given by $_{-2}Y_3^{m}$, and then computed the power-law exponent 
for different values of $m$.
For $m=3$ the initial pulse used for the two-dimensional
evolution corresponds to the lowest possible value of $l$ 
and the late time
power-law behavior is consequently dictated by $\mu = 10.05$.
The situation is different for the case $m=0$, where we find that 
the power-law exponent is given 
by $\mu = 7.72$. That is, the late-time evolution is dominated by the
quadrupole. 
In contrast, the corresponding Schwarzschild evolution
exhibits no such mixing of multipoles, and
evolutions for $l=3,m=0,3$ agree well with the theoretical power-law 
tail exponent $\mu=9$. A result similar to this was discussed in Paper
I.

This result is, however, not too surprising. The mixing of multipoles
is due to the rotational dragging of inertial frames. Basically, there
are two reasons why different
multipoles will be present in the evolution. The first
one is ``imperfect" initial data: Here we have expressed the initial data in
terms of the spin-weighted spherical harmonics  $\ _s Y_l^m$. 
The appropriate angular functions that are used
when the $r$ and $\theta$ components of Eq. (\ref{teuk0}) are separated
are the spin-weighted spheroidal harmonics $\ _s S_l^m(\theta;a\omega)$ \cite{teuk73}).
Hence, because of the frequency (time) dependence in the $\ _s S_l^m(\theta;a\omega)$,
it is difficult to generate initial data for the Kerr problem 
that represent a pure multipole $l$ for a specified value of $m$.
Furthermore, the rotation of the black hole will lead to an active 
coupling between different multipoles (for a discussion of the
analogous
situation for rotating stars, see \cite{Kojstar}). So even if we 
could initiate the evolution with a pure multipole, other
multipoles would be generated and may play a role at later times.

The mixing of multipoles can have an interesting result. Suppose that
the initial data is dominated by a certain multipole $l$ but that 
a relatively small part of $l-1$ is also present. Each of these
multipoles will give rise to power-law tails, but the amplitude of the
latter will be much smaller than the first. In this situation
 the late time field should contain both
 a tail of form $t^{-(2l+3)}$ and another term that behaves as
$t^{-(2l+1)}$. Since the first term dies off faster than the second it
seems unavoidable, no matter how small the amplitude of the second 
tail term is, that the very late time evolution of the field
will be dominated by the $l-1$ multipole.
The
corresponding time evolution will simply exhibit a late-time
change in the angular behaviour.
An example of this was discussed in Paper I.

\subsection{Quasinormal modes}

To further test the performance of the Teukolsky code, we
performed a series of simulations for different values of
the angular momentum parameter $a$ and the
azimuthal number $m$. Then we focussed on the later part of the ringing
sequence, which is dominated by the slowest damped mode of
oscillation, and  read off the oscillation frequency 
$\omega$  via a standard Fourier
transform of the time signal. Table~\ref{qnms} summarizes results
obtained in this way.

A comparison of our values for the real part of each QNM frequency 
with those
given by Leaver \cite{leaver} and Kokkotas \cite{kostas} for $m=0$,
$a/M=0, 0.5, 0.9$ shows an agreement to better than $0.3\%$.
For $m=-1$, our results agree with the values of Detweiler
\cite{detweiler1980} and Leaver (different sign convention)
 to within $0.5\%$. Our results for the imaginary parts are accurate to
$1\%$. 
Although higher multipole evolutions are readily obtained, accurate
estimates of the corresponding QNM  frequencies require
increasingly higher angular resolution. Hence, we have only 
estimated QNM frequencies for the quadrupole modes. But it should be
pointed out that, as was the case for the power-law tails, QNMs 
corresponding to other multipoles are present in all signals. 

Interestingly, we find that the numerically extracted QNM frequencies
for non-zero $m$ do not depend on the sign of $m$,
i.e. we get the same values for the QNM frequencies
from evolutions for e.g. $m =\pm 1$. At
first sight this is surprising, since it seems to contradict well
established results.
According to for example Detweiler \cite{detweiler1980}, the
imaginary parts of the $m=+1$ and $m=-1$ modes should become quite
different as $a$ increases. 

\begin{figure}[tbh]
\leavevmode
\\
\epsfbox{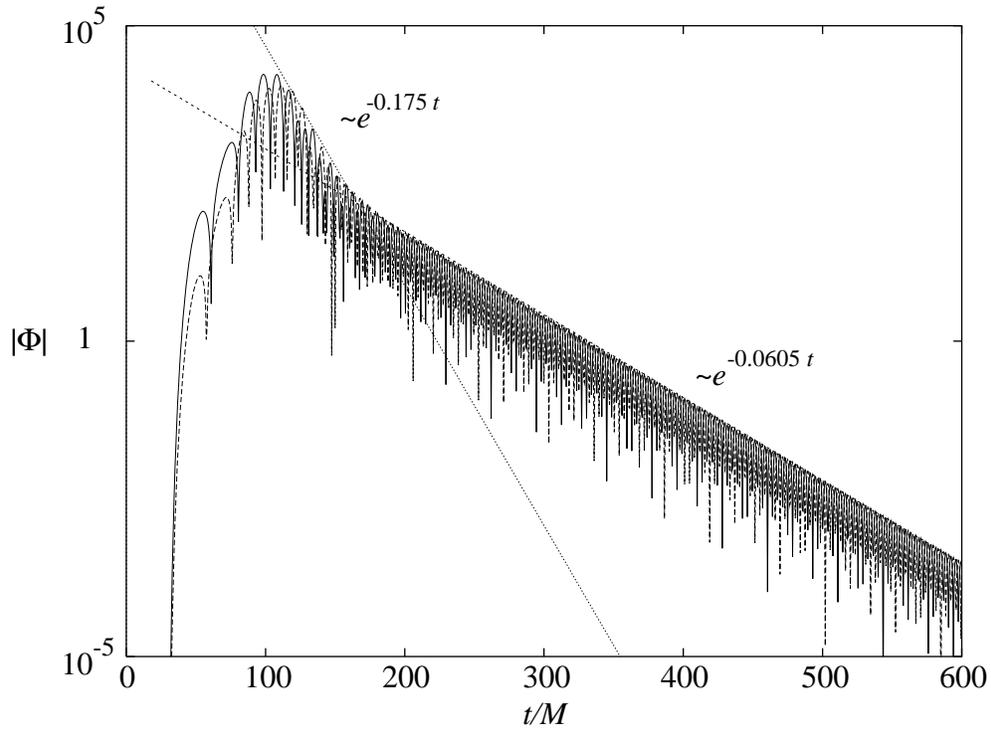}
\caption[figure1]{\label{figure1}
Quasinormal ringing for $a=0.99M$, $M=0.5$, 
and $m=2$. During $120M \leq t \leq
160M$, the field decays as $|\Phi| \propto e^{-0.175 t}$; 
the subsequent evolution is dictated by a slower decaying mode
$|\Phi| \propto e^{-0.0605t}$.
}
\end{figure}

Fortunately, the answer is simple: The frequencies of both 
the $m$ and the $-m$ QNMs 
are present
in a typical evolution. This also 
explains the existence of 
two distinct regimes of the quasinormal ringing that can be seen
in Fig. \ref{figure1}.
In this figure we show the field as seen by an observer at
 $r^* = 20M$, $\theta=\pi/4$ for $a=0.99M$, $M=0.5$, 
and $m=2$. In the early phase of QNM ringing ($120M \leq t \leq
160M$), the decay of the field is approximately given by 
$|\Phi| \propto e^{-0.175 t}$, whereas the behavior for late times 
is governed by a slower decaying mode, $|\Phi| \propto e^{-0.0605t}$. 
These values are in good agreement with the theoretical values for
$l=2$ and $m=\pm2$, as can be seen by comparison with Fig. 1(a) in 
\cite{detweiler1980}.

One question remains. Why should we expect both these modes to be
present in the signal? The answer can be found in the symmetries of the
problem.  In the Schwarzschild case we know that 
there  will be modes in the first and second quadrant 
of the complex $\omega$-plane (assuming that the modes have a time-dependence
$e^{i\omega t}$). All these modes will contribute to the 
signal, but since they occur as complex conjugate pairs $\omega_l$ and
$-\omega_l^\ast$  (where the asterisk represents
complex conjugation) the contribution from one set of
modes can be expressed in terms of the other. Hence, the QNM 
part of the signal can be evaluated using only modes in the first
quadrant (for a study of the Schwarzschild problem,
see \cite{Nils}).

In the case of Kerr black holes the situation is more complicated.
The QNMs no longer occur as complex
conjugate pairs. Let us suppose that we work in the frequency domain, and
that we have a solution 
$\Psi_{lm}(\omega, r, \theta)e^{im\phi}$  to the Teukolsky equation for  given $l$ and $m$. Then it is easy to show
(using the separated
equations, see \cite{teuk73}) that a solution to the equation for $-m$ will 
follow as
\begin{equation}
\Psi_{l-m}(\omega, r, \theta) = [ \Psi_{lm} (-\omega^\ast, r, \pi - \theta) ]^\ast \ .
\end{equation} 
This means that
if $\omega$ is a
QNM corresponding to $l$ and $m$ then the complex conjugate
$-\omega^\ast$ will be a QNM for 
$l$ and $-m$. This symmetry is nicely illustrated in Fig. 3 in
\cite{leaver}. 
We can use this symmetry to express the contribution
from the Kerr QNMs
in the second quadrant in terms of modes in the first quadrant
(we will describe this procedure in detail elsewhere). If we represent
modes in the first quadrant by $\omega_{lmn}$, we can   
schematically write the QNM contribution to the signal as
\begin{equation}
\Psi_m^{QNM} = \sum_{ln} \left\{ F(\omega_{lmn}) + G(-\omega_{l-mn}^\ast)
\right\} \ .
\end{equation}
This explains the presence of both branches of QNMs in our evolutions.

\subsection{Wave propagation}

Finally, an interesting direct application of our code
is the identification of effects of the
background rotation on the propagation of signals.
The idea would be to use the same initial data for black holes with different
rotation rates and see to what extent the rotation of the black hole
can be inferred from the emerging signals.
A stumbling block for work in this direction, though, is the
difficulty to prescribe initial data that represent the
``same'' initial physical perturbation for any value of the
black hole angular momentum.
This prescription is possible for data coming in from
infinity, since the effects of rotation are negligible for large enough
radii. However, it is not clear how to set up initial data
(e.g. position of the pulse)
in the vicinity of the black hole in such a way that comparison
of evolutions for different angular momenta makes sense.
For this reason, we addressed only the effect of rotation on
 pulses that were initially far way from the hole.

\begin{figure}[tbh]
\leavevmode
\\
\epsfxsize=0.8\textwidth
\epsfbox{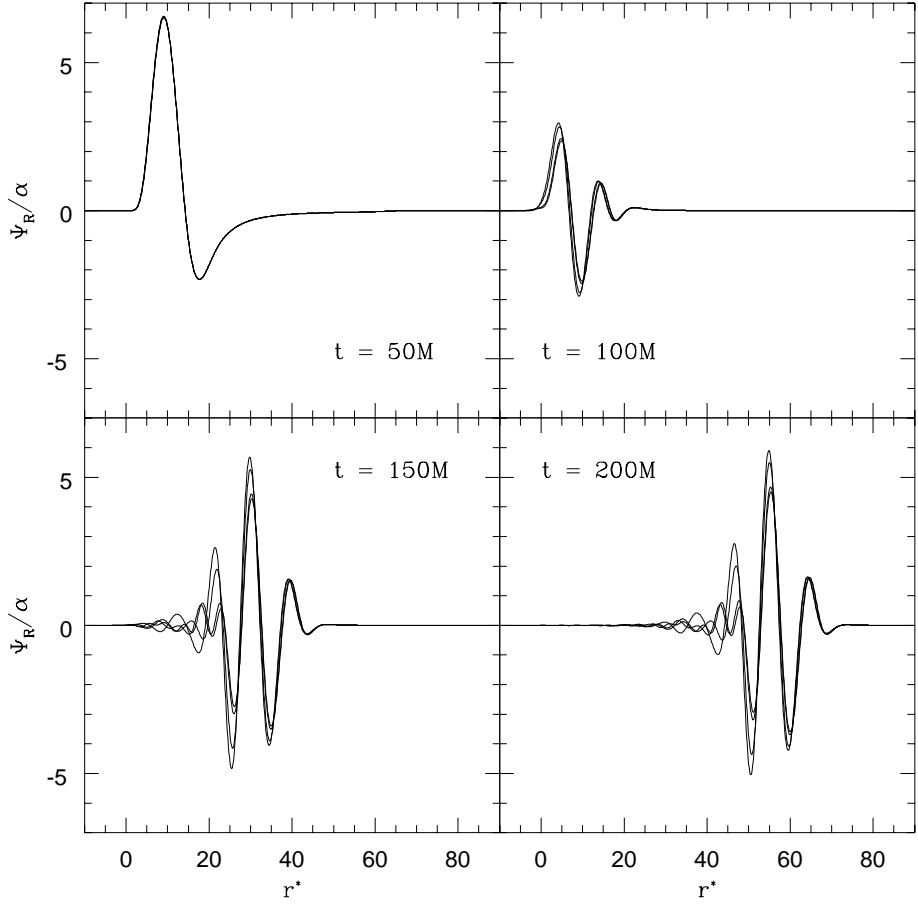}
\caption[figure2]{\label{figure2}
Snapshots of the evolution of $\Phi_R$ for $m=2$, at $\theta=\pi/2$, in the
time interval $0\leq t \leq 200M$.
For clarity, $\Phi_R/\alpha$ is displayed with $\alpha$ a scale factor:
$\alpha = (10,\,10^3,\, 10^3,\, 10^3)$ for 
$t = (50,\, 100,\, 150,\, 200)\,M$, respectively.
The values used for the angular momentum are $a/M=0,\, 0.5,\, 0.9,\, 0.99$.
The initial data consist of a bell-shaped pulse 
centered at $75M$ and propagating inwards.
Differences in the pulses do not become appreciable until
the pulse has been reflected at the potential barrier outside the black hole
(see $t = 100M$ snapshot).
By $t= 200M$ a decrease of $\sim 25\%$ in the wavelength can be observed.
The maximum amplitude of the pulse also  decreases with $a$.
}
\end{figure}

In Fig. \ref{figure2} we show a series of snapshots in $r^*$ 
of the evolution of a pulse for $m=2$. 
At $t=0$, the initial data consists of a bell-shaped pulse in $r^*$ 
centered at $75M$ and propagating inwards.
Then  the 
evolution of $\Phi_R$ is shown at constant $\theta=\pi/2$ and $0\leq t \leq
200M$  for four different values of 
the angular momentum of the black hole: $a/M=0,\, 0.5,\,
0.9,\, 0.99$.  
One immediately sees that an increase in the 
angular momentum of the black hole has two effects on the
scattered pulse. Both the amplitude and
wavelength of the emerging signal change. 
By $t= 200M$ the wavelength in the trailing part of the signal 
for $a/M=0.99$ is approximately half as long as for the 
non-rotating case. This agrees well with the 
anticipated increase in the dominating QNM frequency \cite{detweiler1980}.
Furthermore,  at $t=200M$ the amplitudes for
$a/M=0$ and $a/M=0.99$ also differ by $\sim 25\%$.
The amplitude of the emerging signal generally decreases as we increase
$a$. 

\begin{figure}[tbh]
\leavevmode
\\
\epsfxsize=0.8\textwidth
\epsfbox{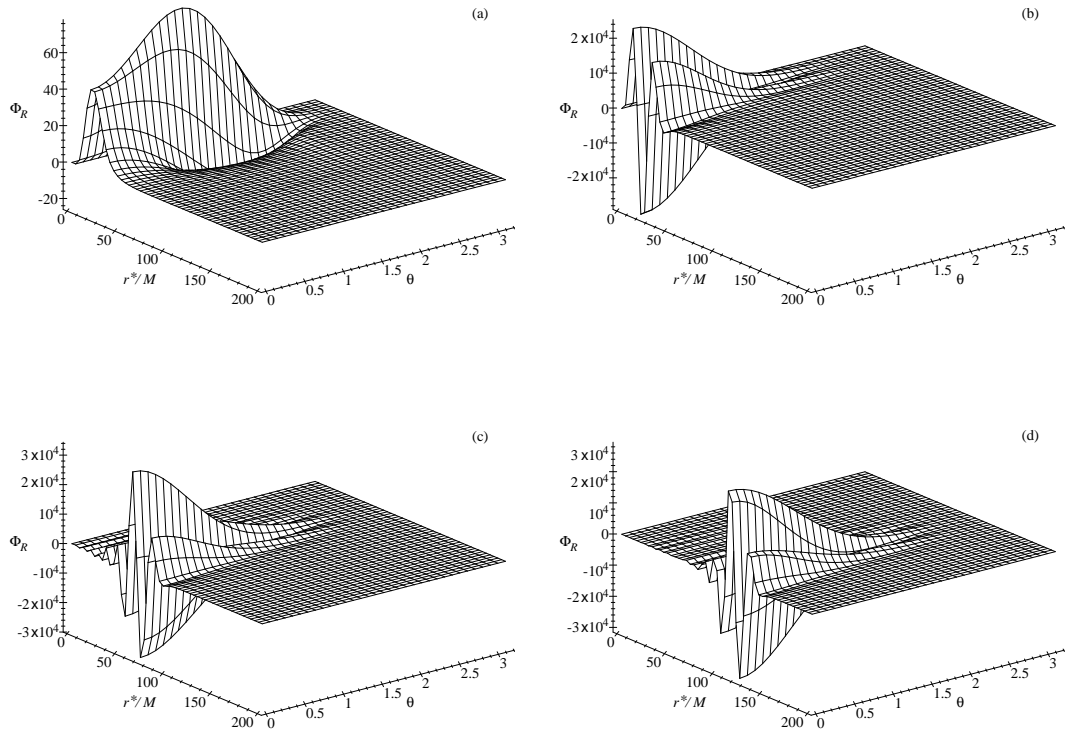}
\caption[figure3]{\label{figure3} Evolution for 
    $a/M = 0.99$ at $t=50M$ (a), $t=100M$ (b), $t=150M$ (c), and $t=200M$ (d). 
    The initial data consist of a bell-shaped pulse centered at $r^*=75M$,
    $\theta=\pi/2$
    and propagating inwards. During the evolution there is a clear
    transition 
    of the
    angular distribution to $_{-2}Y_2^2 \propto \cos^4\theta/2$.
                 }
\end{figure}

The angular behaviour for the scattering scenario also shows interesting
features. An example can be seen in
Fig. \ref{figure3} where we show the evolution for 
$a/M = 0.99$.
The initial data is given by $\Phi_R\propto \sin^2\theta$, 
centered in $r^*$ at $75M$ and propagating inwards, but after some time
there is a clear transition of the
angular distribution to $_{-2}Y_2^2 \propto \cos^4\theta/2$. This
is the dominant angular behavior during the later
evolution. That is,  the quadrupole dominates the scattered
wave at later times. This situation is similar to the one we discussed for power-law tails.

\section{Conclusions}

We have studied the time evolution of perturbations of
rotating black holes by numerically integrating the
Teukolsky equation written as a system of equations
of first-order in time and space.
In particular, we investigated the role of the
black holes angular momentum on the dynamics during the burst, 
quasinormal ringing and power-law
tail phases. We have reproduced values for the fundamental
quasinormal mode frequencies with an accuracy of better than $1\%$.
As for the late-time regime, we found power-law tail behavior analogous
to the scalar field case that we had studied previously.
These results are in good agreement with what one would expect
and establish the reliability and accuracy of our code. We thus
have access to a useful numerical laboratory that can be used to
further illuminate the detailed physics of rotating black holes. 

In the future we plan to use this laboratory to investigate several interesting issues. We would, for example, like to establish what effect
the so-called super-radiance will have on the evolution of initial data. 
From studies of scattering of monochromatic waves from rotating black holes
(see, for example \cite{Matzner}) it is know  that frequencies such
that $\omega M < am/2r_+$ will correspond to a reflection coefficient whose
magnitude is larger than unity. That is, these frequencies are ``super-radiant''.
This is the wave-analogue to the well-known Penrose process. 
In the results presented in this paper the effect of super-radiance
was not obvious, but it is 
plausible that it can be unveiled in a more detailed study. For example, 
it seems that super-radiance should lead to different results
for co- and counter-rotating waves in an evolution. That is, if super-radiance
is relevant one should see difference between evolutions for $|m|$ and $-|m|$.

Another interesting issue concerns the excitation of quasinormal modes.
It is known that the damping of the $-|m|$ modes will vanish as $a$
increases \cite{leaver}. This would potentially make the modes
of a rapidly rotating black hole ideal for gravitational wave
detection.
However, in an approximate study Ferrari and Mashhoon \cite{Ferrari} 
have shown that
the amplitude of these extremely slowly damped modes will be
negligible. In the 
extreme Kerr limit no energy is expected to be radiated through these modes.
This result has proved to be very hard to verify analytically, but
it is testable with the present code. We should thus be able to establish whether 
the extremely slowly damped quasinormal modes are of astrophysical relevance or not. 

Another outstanding problem in black-hole physics concerns the dynamical 
stability of the Kerr black hole to perturbations. So far, only mode-stability
has been established \cite{whiting}. To prove complete stability one must ensure that
all relevant quantities remain point-wise bounded during an evolution, cf. \cite{kay}. It seems plausible that our Teukolsky code can prove useful also in this context. 
 
Perhaps the most interesting future 
application of our numerical Teukolsky code 
will be to evolve initial data qualitative
similar to the late stages of a binary black
hole coalescence. 
The main problem in approximating black hole collisions
with perturbations about Kerr spacetimes is the construction of initial data.
Work so far on the close limit approximation to construct initial data
has heavily depended on having a conformally flat background geometry.
On the other hand, constant time hypersurfaces in
Boyer-Lindquist or Kerr coordinates are not conformally flat. Hence
disentangling these effects is not as straightforward as for nonrotating black holes.
Possible ways of circumventing these obstacles are
presently being investigated \cite{ThePaper}.

\section{Acknowledgments}

We thank  K.\ Kokkotas, H.-P.\ Nollert, R.\ Price and J.\ Pullin for
helpful discussions. This work was supported by the Binary Black Hole
Grand Challenge Alliance, NSF PHY/ASC 9318152 (ARPA supplemented) and
by NSF grants PHY 96-01413, 93-57219 (NYI) to PL. WK was
supported by the Deutscher Akademischer Austauschdienst (DAAD) and the
Deutsche Forschungsgemeinschaft (DFG), SFB 382.
NA is supported by NSF grant PHY 92-2290. 

\appendix

\section{}\label{appendix}
In the following, we give the explicit form of the coefficients
in the matrices {\boldmath{$M$}}, {\boldmath{$A$}}, and {\boldmath{$L$}}
in Eq.\ (\ref{new_teu}).
With
\begin{eqnarray}
c_1 & \equiv&
2\,s\,\frac {-3M{r}^{2}+M{a}^{2}+{r}^{3}+r{a}^{2}}
{\Sigma^2} \; , \\
c_2 &\equiv&
-2\,\frac{r\,\Delta\,(1+s)-M\,(a^2-r^2)\,s}
{\Sigma^2} \; , \\
c_3 &\equiv&
2\,a\,{\frac
{2\,Mrm+\Delta\,s\,\cos\theta}
{\Sigma^2}}\; , \\
c_4 &\equiv&
-2\, a \, m {\frac
{{r}^{2}+{a}^2}{
\Sigma^2
}} \; , \\
\end{eqnarray}
the coefficients of {\boldmath{$M$}} can be written as
\begin{mathletters}
\begin{eqnarray}
m_{31} &\equiv& -b \,c_1 +b\,\frac{\partial b}{\partial r^*}+c_2 \\
m_{32} &\equiv& b \,c_3 - c_4 \; .
\end{eqnarray}
\end{mathletters}
Defining
\begin{eqnarray}
c_5 &\equiv&
\frac{\Delta (  - {m}^{2}
- 2\,\cos\theta \,s\,m
-\cos^2 \theta\,{s}^{2}
+\sin^2\theta \,s) }
{\Sigma^2 \sin^2 \theta
}
 \; , \\
c_6 &\equiv&
-4\,{\frac {\left (r-M\right
)\,s\,m\,a}{\Sigma^2}}
 \; , \\
\end{eqnarray}
the coefficients of {\boldmath{$A$}} are given by
\begin{mathletters}as well as for
investigating asymptotic behavior
\begin{eqnarray}
a_{31} &\equiv& c_5 \; , \\
a_{32} &\equiv& -c_6 \; , \\
a_{33} &\equiv& c_1 \; , \\
a_{34} &\equiv& - c_2 \; . \\
\end{eqnarray}
\end{mathletters}
Finally, with
\begin{eqnarray}
c_7 &\equiv&  -{\frac
{\Delta}{\Sigma^2}} \; , \\
c_8 &\equiv&
-\cot \theta\,{\frac {\Delta}
{\Sigma^2}}
 \; ,
\end{eqnarray}
the only non-vanishing coefficient of the operator matrix
{\boldmath{$L$}} reads
\begin{eqnarray}
l_{31} &\equiv&  c_7 \,\frac{\partial^2}{\partial \theta^2}
           +c_8 \, \frac{\partial}{\partial \theta} \; .
\end{eqnarray}

\begin{table}[tailvalues]\caption{\label{tailvalues}
The numerically determined power-law exponents $\mu$ governing the
late-time behavior of $|\Phi| \propto t^{-\mu}$ measured at $r^*=20M$,
$\theta=\pi/2$. The theoretical value for $a=0$ is $\mu=7$.
}
\begin{tabular}{c|c|c}
 $a/M$    &    $\mu \, (m=0)$ & $\mu \, (m=1)$ \\ \hline
  0     &         7.70   &        7.69         \\
  0.25  &         7.73   &        7.77         \\
  0.50  &         7.75   &        7.68         \\
  0.75  &         7.81   &        7.79         \\
  0.90  &         7.68   &        7.71         \\
\end{tabular}
\end{table}

\begin{table}[qnms]\caption{\label{qnms}
The numerically extracted QNM frequencies 
that dominate the late-time part of the ringing phase for $m=0,\pm 1$.
}
\begin{tabular}{c|c|c}
 $a/M$    &    $\omega M \, (m=0)$ & $\omega M \, (m=|1|)$  \\ \hline
  0     &      $  0.373 + i\, 0.0885$  &   $0.373+i\,0.0885$   \\
  0.25  &      $0.376+i\,0.0885$  &    $0.392+i\,0.0885$   \\
  0.50  &      $0.383+i\,0.0865$  &    $0.420+i\,0.0860$   \\
  0.75  &      $0.398+i\,0.0835$  &    $0.467+i\,0.0795$   \\
  0.90  &      $0.412+i\,0.0780$   &   $0.517+i\,0.0695$   \\
\end{tabular}
\end{table}

%
%
%
%
%
%
%
\end{document}